\documentclass{iaus}
\usepackage{graphicx}

\title[Star Formation in the Eagle Nebul] 
{Star Formation in the Eagle Nebula and NGC\,6611}

\author[Oliveira, Jeffries \& van Loon]   
{J.M. Oliveira, R.D. Jeffries \& J.Th van Loon}

\affiliation{Astrophysics Group, Lennard-Jones Laboratories, Keele University, 
UK}

\pubyear{2006}
\volume{237}  
\pagerange{1--1}
\date{?? and in revised form ??}
\jname{Triggered Star Formation in a Turbulent ISM}
\editors{B. G. Elmegreen \& J. Palous, eds.}
\begin{document}

\maketitle

\begin{abstract}
We present $IZJHKL'$ photometry of the core of the cluster NGC\,6611 in the
Eagle Nebula. This photometry is used to constrain the Initial Mass Function
(IMF) and the circumstellar disk frequency of the young stellar objects. Optical
spectroscopy of 258 objects is used to confirm membership and constrain
contamination as well as individual reddening estimates. Our overall aim is to
assess the influence of the ionizing radiation from the massive stars on the
formation and evolution of young low-mass stars and their disks. The disk
frequency determined from the $JHKL'$ colour-colour diagram suggests that the
ionizing radiation from the massive stars has little effect on disk evolution
(\cite[Oliveira et al. 2005]{oliveira05}). The cluster IMF seems 
indistinguishable from those of quieter environments; however towards lower
masses the tell-tale signs of an environmental influence are expected to become
more noticeable, a question we are currently addressing with our recently
acquired ultra-deep (ACS and NICMOS) HST images.
\end{abstract}

\begin{figure}[h]
 \centerline{
 \scalebox{0.66}{\includegraphics{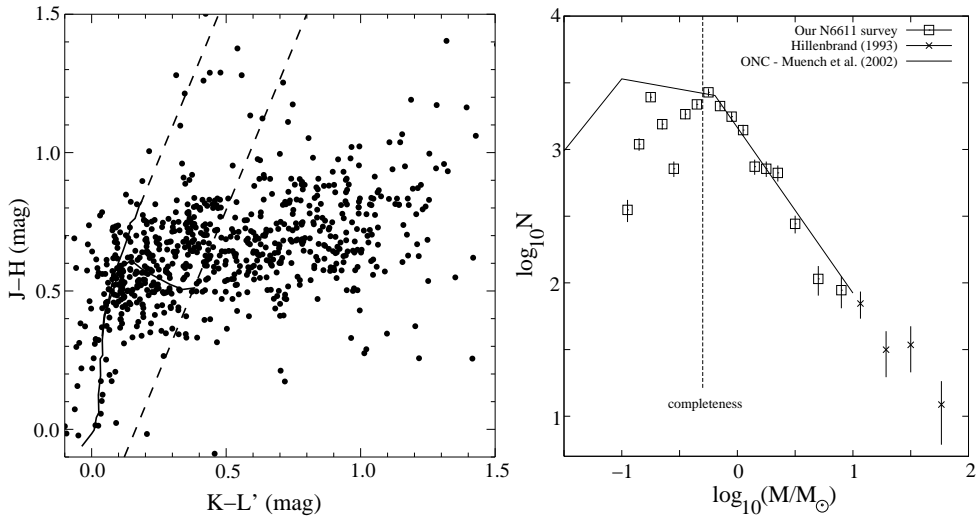}}}
  \caption{$JHKL'$ colour-colour diagram and IMF for NGC\,6611. Down to objects
  of $\sim$\,0.5\,M$_{\odot}$, neither the disk frequency nor the IMF show
  significant evidence for a strong ionizing influence from the massive O-stars
  in the cluster (\cite[Oliveira et al. 2005]{oliveira05}).}
\end{figure}

\end{document}